\documentclass[12pt]{iopart}
\usepackage{iopams}
\usepackage{epsfig}
\usepackage{textcomp}
\usepackage{citesort}
\begin{document}

\title[Self-organization of punishment in structured populations]{Self-organization of punishment in structured populations}

\author{Matja{\v z} Perc$^1$ and Attila Szolnoki$^2$}
\address{$^1$Faculty of Natural Sciences and Mathematics, University of Maribor, Koro{\v s}ka cesta 160, SI-2000 Maribor, Slovenia\\
$^2$Institute of Technical Physics and Materials Science, Research Centre for Natural
Sciences, Hungarian Academy of Sciences, P.O. Box 49, H-1525 Budapest, Hungary}
\ead{matjaz.perc@uni-mb.si, szolnoki@mfa.kfki.hu}

\begin{abstract}
Cooperation is crucial for the remarkable evolutionary success of the human species. Not surprisingly, some individuals are willing to bare additional costs in order to punish defectors. Current models assume that, once set, the fine and cost of punishment do not change over time. Here we show that relaxing this assumption by allowing players to adapt their sanctioning efforts in dependence on the success of cooperation can explain both, the spontaneous emergence of punishment, as well as its ability to deter defectors and those unwilling to punish them with globally negligible investments. By means of phase diagrams and the analysis of emerging spatial patterns, we demonstrate that adaptive punishment promotes public cooperation either through the invigoration of spatial reciprocity, the prevention of the emergence of cyclic dominance, or through the provision of competitive advantages to those that sanction antisocial behavior. Presented results indicate that the process of self-organization significantly elevates the effectiveness of punishment, and they reveal new mechanisms by means of which this fascinating and widespread social behavior could have evolved.
\end{abstract}

\maketitle

\section{Introduction}
It is challenging to arrive at a satisfactory evolutionary explanation for the observed collective efforts in large groups consisting of genetically unrelated individuals
\cite{fehr_aer00}. Those that exploit the public goods but do not contribute to them obviously fare better, and should thus be able to outperform cooperators in an environment where the more successful strategies are more likely to spread. Along with the rise of defectors, however, comes the inevitable failure to harvest the benefits of a collective investment, and the society may evolve towards the ``tragedy of the commons'' \cite{hardin_g_s68}. Yet humans have mastered the act of large-scale cooperation almost to perfection \cite{sigmund_10,nowak_11}, which is something that distinguishes us markedly from other species \cite{hrdy_11}.

Besides traditional kin and group selection as well as different forms of reciprocity \cite{nowak_s06}, complex interaction networks \cite{lozano_ploso08,gomez-gardenes_c11,gomez-gardenes_epl11}, diversity \cite{santos_n08,wang_j_pre10b,perc_njp11,santos_jtb12}, risk of collective failures \cite{santos_pnas11}, selection pressure \cite{van-segbroeck_njp11}, reputation-based partner choice \cite{fu_pre08b,wu_t_pre09}, conditional strategies \cite{szolnoki_pre12}, as well as both the joker \cite{arenas_jtb11} and the Matthew effect \cite{perc_pre11}, there exist compelling theoretical and empirical evidence pointing towards punishment as a successful means to ensure high levels of cooperation in groups whose members are driven by selfishness \cite{sigmund_tee07}. Recent research highlights both individually inspired actions (peer-punishment) \cite{fehr_n02,gardner_a_an04,brandt_pnas06,gachter_s08,rockenbach_n09} as well as sanctioning institutions (pool-punishment) \cite{gurerk_s06,sigmund_n10,szolnoki_pre11b} as viable for promoting the evolution of cooperation. However, the execution of punishment may be costly, which weighs heavily on the shoulders of those that already contribute to the public good \cite{fehr_n03,fowler_pnas05,milinski_n08}. Whether or not to sanction defectors therefore becomes a similar dilemma as whether to contribute to the public good or not \cite{fehr_n04}. Even if punishment can be sustained, its costs often exceed the benefits of enhanced cooperation. Reputation \cite{panchanathan_n04}, the possibility to abstain from the joint enterprise \cite{hauert_s07}, as well as coordinated efforts between the punishers \cite{boyd_s10}, have all been identified as potential solutions to these problems.

While pioneering game theoretical approaches entailing punishment focused on analytically tractable models where every player can interact with everybody else, there also exist works where more realistic structured populations with limited interaction were considered \cite{brandt_prsb03,nakamaru_eer05,helbing_ploscb10,szolnoki_pre11}. Structure may offer new solutions via pattern formation that is brought about by several competing strategies. This includes the spontaneous formation of alliances, the emergence of cycling dominance, as well as logarithmically slow coarsening \cite{helbing_pre10c}, to name but a few. Taking into account structure within a population has indeed lead to interesting and sometimes counterintuitive results in terms of the effectiveness and sustenance of punishment. Previously studied models, however, cannot account for the fact that even punishers may be reluctant to bare, or at least they will try to minimize, additional costs when the threat of punishment is either no longer necessary or proves to be ineffective. Similarly, models where sanctioning is considered as an unchanging part of ones strategy fail to acknowledge a common real-life observation, which is that an increase in antisocial behavior will frequently trigger an increase in both willingness as well as severity of sanctioning amongst those who feel threatened by the negative consequences. Extreme examples thereof include terrorist attacks and other malicious acts, upon which the security measures in the affected areas are often tightened rather drastically.

In this paper, we take these arguments into account and extend the existing theory on sanctioning in structured populations by introducing and studying adaptive punishment in the spatial public goods game. Our model initially contains only cooperators and defectors as the two strategies competing for space on the square lattice. However, cooperators may resort to punishing defectors if the latter spread successfully. Whenever a defector succeeds in passing its strategy to a cooperator, all the remaining cooperators who witness this will increase the fine that will be inflicted on the neighboring defectors by a factor $\Delta$. It is important to note that cooperators will not necessarily start or increase the severity of punishing in the presence of defectors. In fact, they will do so only if they observe a negative tendency in the evolution of cooperation, \textit{i.e.} the successful spreading of defectors. Upon increasing the punishment fine, however, the cost cooperators will have to bare also increases accordingly, but it is scaled by a free parameter $\alpha$ that determines just how costly the sanctions are. Accounting for the permanent strive of players to maximize their fitness is a corresponding decrease in fine and cost that applies to all punishing cooperators after a full round of the game. Thus, there is a constant drift towards non-punishment as long as defectors do not spread. Making full use of evolutionary competition, strategy invasions are possible not just between cooperators and defectors, but also between pure cooperators and cooperators with different punishing activity. In this way pure cooperation can be selected via imitation and cooperators that do not punish are free to overtake those that do punish with an effective fine. Irrespective of their punishing activity, however, all cooperators contribute a fixed amount to the public good. The income that accumulates is multiplied by the synergy factor $r$ and then divided equally amongst all the group members irrespective of their strategy and punishing activity.

As we will show in what follows, this model is able to explain the spontaneous emergence of punishers and is very effective in sustaining collective efforts in the population, while at the same time leading to an evolutionary dynamics where the footprint in terms of resources spent on sanctioning is negligible. This is achieved by means of newly identified mechanisms of pattern formation that cannot emerge in the absence of self-organization.

\section{Spatial public goods game with adaptive punishment}\label{Model}
The game is staged on a square lattice with periodic boundary conditions. Players play the game with their $k=4$ nearest neighbors. Accordingly, each individual belongs to $G=5$ different groups containing $k+1$ players each. Initially each player on site $x$ is designated either as a cooperator ($s_x = {\rm C}$) or defector ($s_x = {\rm D}$) with equal probability. Using standard parametrization, cooperators contribute $1$ to the public good while defectors contribute nothing. The sum of all contributions in each group is multiplied by the factor $r>1$, reflecting the synergetic effects of cooperation, and the resulting amount is equally divided amongst the $k+1$ members irrespective of their strategy.

To accommodate adaptive punishment, each player is assigned an additional parameter $\pi_x$ keeping score of its punishing activity. Initially $\pi_x=0$ for all players. Subsequently, whenever a defector succeeds in passing its strategy, all the remaining cooperators in all the groups containing the defeated cooperator increase their punishing activity by one, \textit{i.e.} $\pi_x=\pi_x+1$. We emphasize that the presence of defectors alone never triggers an increase in $\pi_x$. Only when defectors spread do the cooperators resort to sanctioning. Conversely, if defectors fail to spread, then at every second round all cooperators decrease their punishing activity by one, as long as $\pi_x \geq 0$. Note that if the population contains cooperators with $\pi_x>0$ then all the cooperators having $\pi_x=0$ become second-order free-riders. Due to the presence of punishers, \textit{i.e.} cooperators having $\pi_x>0$, the accumulation of payoffs changes as well. In particular, each defector is fined with an amount $\pi_x \Delta/k$ from every punishing cooperator that is a member of the group, while at the same time the punishing cooperators that execute the punishment bare the cost $\pi_x \alpha \Delta/k$ for every defector punished. Here $\Delta$ determines the incremental step used for the punishing activity and $\alpha$ is a free parameter determining whether the sanctions are costly ($\alpha>1$) or not ($\alpha<1$). Taking punishment into account, the payoff of player $x$ in a given group $g$ is thus
\begin{equation}
P_{\rm C}^g=r \frac{N_{\rm C}}{G} - 1 - \frac{1}{k}N_{\rm D} \pi_x \alpha \Delta \,\,\,\,\,\,\,\,\,\, {\rm if} \,\,\,\, s_x={\rm C} \,\, ,
\label{PC}
\end{equation}
and
\begin{equation}
P_{\rm D}^g=r \frac{N_{\rm C}}{G} - \frac{1}{k} \sum_{y \in g} \pi_y \Delta \,\,\,\,\,\,\,\,\,\,\,\,\,\,\,\,\,\,\,\,\,\, {\rm if} \,\,\,\, s_x={\rm D} \,\, ,
\label{PD}
\end{equation}
where $N_{\rm C}$ and $N_{\rm D}$ are the numbers of cooperators and defectors in the group $g$, respectively.

The stationary fractions of cooperators $\rho_{\rm C}$ and defectors $\rho_{\rm D}$ on the square lattice are determined by means of a random sequential update comprising the following elementary steps. First, a randomly selected player $x$ plays the public goods game with its $k$ interaction partners as a member of all the $g=1, \ldots, G$ groups it belongs to. The overall payoff it thereby obtains is thus $P_{s_x} = \sum_g P_{s_x}^g$. Next, one of the four nearest neighbors of player $x$ is chosen randomly, and its location is denoted by $y$. Player $y$ also acquires its payoff $P_{s_y}$ identically as previously player $x$. Finally, if $s_x \neq s_y$ player $y$ imitates the strategy of player $x$ with the probability $q=1/\{1+\exp[(P_{s_y}-P_{s_x})/K]\}$, and in case of successful imitation player $y$ resets its punishing activity to zero ($\pi_y=0$). If, however, $s_x = s_y = {\rm C}$ player $y$ adopts the punishing activity $\pi_x$ from player $x$ with the same probability $q$, while if $s_x = s_y = {\rm D}$ nothing happens. Here $K$ determines the level of uncertainty by strategy adoptions or its inverse $K^{-1}$ the so-called intensity of selection. Without loss of generality we set $K=0.5$ \cite{szolnoki_pre09c}, implying that better performing players are readily imitated, but it is not impossible to adopt the strategy (or the punishing activity) of a player performing worse. Such errors in judgment can be attributed to mistakes and external influences that affect the evaluation of the opponent. Each full round (Monte Carlo step) of the game involves all players having a chance to adopt a strategy from one of their neighbors once on average. Depending on the proximity to phase transition points and the typical size of emerging spatial patterns, the linear system size was varied from $L=200$ to $7000$ and the equilibration required up to $10^7$ full rounds of the game for the finite size effects to be avoided.

This set-up also enables us to directly compare the effectiveness of the self-organization of punishment with steady punishment efforts studied previously in \cite{helbing_ploscb10}. While the simulation details are identical in both cases, in the steady punishment model initially some portion of the square lattice is populated by punishing cooperators who punish every defector with a fine $\Delta/k$ and therefore bare the cost of sanctioning $\alpha \Delta /k$. The initially set punishing activity of punishing cooperators $\pi_x=1$ never increases or decreases, and if they succeed in spreading the new punishing cooperators have $\pi_x=1$, while if being overtaken by either non-punishing cooperators or defectors they altogether loose their punishing ability. For further details we refer to \cite{helbing_ploscb10}, where the steady punishment model was presented and studied in detail.

\section{Results}\label{Results}

\subsection{Phase diagrams}

\begin{figure}
\centerline{\epsfig{file=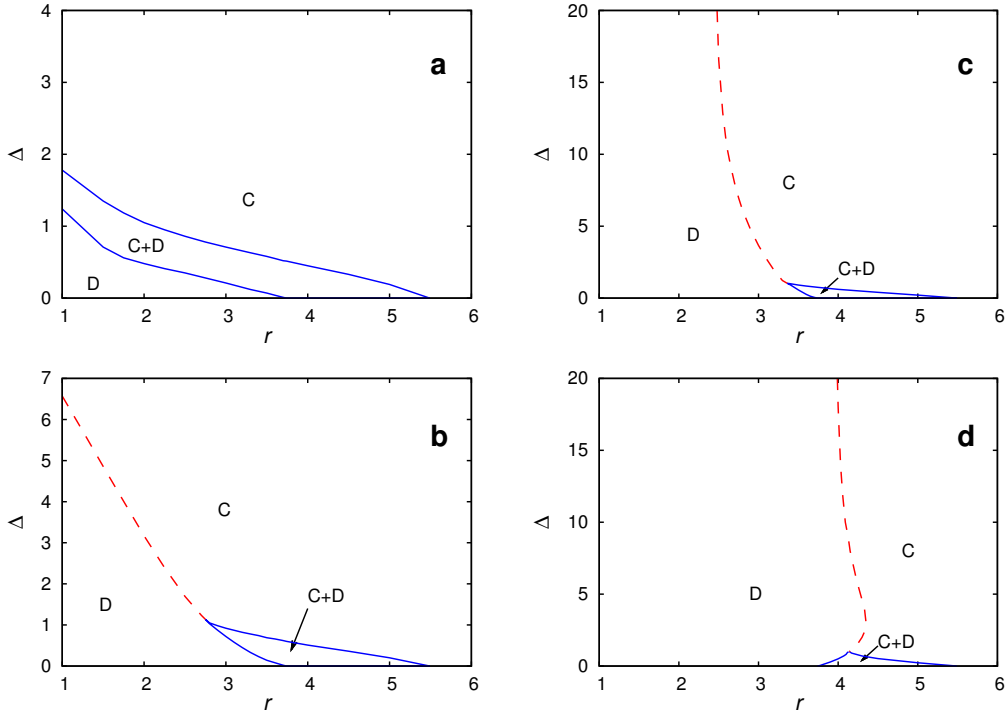,width=14cm}}
\caption{Phase diagrams for the spatial public goods game with adaptive punishment. Depicted are the strategies (C - cooperators, D - defectors) that remain on the square lattice after sufficiently long relaxation times in dependence on the synergy factor $r$ and the incremental step used for adapting the punishing activity $\Delta$. The displayed phase diagrams are for (a) $\alpha=0.1$, (b) $\alpha=0.5$, (c) $\alpha=1$ and (d) $\alpha=2$. (a) If sanctioning is inexpensive (note that $\alpha=0.1$ implies that punishment costs are only $1/10$ of the fines imposed on defectors), already small values of $\Delta$ suffice to restore a mixed C+D phase or even a pure C phase at synergy factors where otherwise defectors would reign supreme. Continuous phase transitions, depicted by solid blue lines, are the only means through which the stability of the two strategies changes. (b) Raising the cost of punishment to $\alpha=0.5$ leads to a qualitative change in the evolutionary dynamics. Below $r=2.7$ the mixed C+D phase is no longer possible. Instead, discontinuous phase transitions, depicted by the dashed red line, lead to the complete dominance of cooperators for sufficiently high values of $\Delta$. Expectedly, the lower the synergy factor $r$, the higher the value of $\Delta$ required to reach the pure C phase. Still, even at $r=1$ cooperators are able to dominate completely if $\Delta>6.5$. (c) The $\alpha=1$ case represents the border between inexpensive and costly punishment, implying that the fine and cost of punishment are equal. While qualitatively the evolutionary dynamics does not change if compared to $\alpha=0.5$, we first observe a hard limit in $r$, below which cooperation can no longer be sustained, irrespective of how large $\Delta$ is. For $\alpha=1$ we find the limiting synergy factor to be $r=2.48$. (d) Increasing the cost of sanctioning further to $\alpha=2$ increases the minimal value of $r$ where cooperators are able to survive. While the mixed C+D phase is possible by means of a continuous phase transition at as low as $r=3.74$, the pure C phase is unattainable if $r<3.99$.}
\label{phase}
\end{figure}

In Fig.~\ref{phase}, we present full $r-\Delta$ phase diagrams for different values of $\alpha$. These feature the survivability of the two competing strategies on the square lattice after relaxation, when the stationary state is reached. Depending on the parameter values, cooperators (C) and defectors (D) can dominate completely, although a coexistence of the two strategies (C+D) is possible as well. In all panels solid blue lines depict continuous (second-order) phase transitions, while dashed red lines depict discontinuous (first-order) phase transitions. If the punishment is not costly (panels a and b), the social dilemma can be resolved completely, and accordingly, full cooperator dominance is always possible. That is to say, there exists a sufficiently large value of $\Delta$ irrespective of $r$, such that defectors are unable to spread and are eventually completely eliminated from the population.

If the costs of sanctioning relative to the imposed fines are negligible (panel a), the pure C phase emerges by means of a continuous phase transition from the mixed C+D phase, while for somewhat larger $\alpha$ (panel b) the outbreak of cooperation is possible also by means of a discontinuous phase transition. As the sanctioning becomes costly (panels c and d), however, the limits of adaptive punishment become clearly inferable. With increasing values of $\alpha$, both the survivability (mixed C+D phase) as well as the potential dominance (pure C phase) of cooperators shift towards larger values of $r$, whereby as by smaller values of $\alpha$, both continuous as well as discontinuous phase transitions can be observed.

In the absence of punishment cooperators survive only if $r>3.74$, and are able to crowd out defectors completely for $r>5.49$ \cite{szolnoki_pre09c}. Taking these as benchmark values for evaluating the impact of adaptive punishment as presented in Fig.~\ref{phase}, we conclude that it provides ample support for the evolution of cooperation. Expectedly, the support is more effective if the cost of sanctioning is small, but even if the punishment is costly, the evolutionary advantage for public cooperation granted by adaptive punishment is undeniable.

It is also instructive to compare the effects of the self-organization of punishment with those reported previously for steady punishment efforts, \textit{i.e.} when the punishers are introduced as a separate strategy and use a constant fine to sanction defectors \cite{helbing_ploscb10,brandt_prsb03,nakamaru_eer05}. Although steady punishment efforts on structured populations have been found effective for the promotion of cooperation, for example by means of the segregation of punishing cooperators and second-order free-riders, or by means of the spontaneous emergence of alliances between different strategies, a more detailed analysis reveals that adaptive punishment may provide even stronger and so far unknown incentives for collaborative efforts in sizable groups. In what follows, we will examine the evolution of characteristic spatial patterns formed by the competing strategies that emerge in different regions of the presented phase diagrams (Fig.~\ref{phase}). For reference and comparison purposes, we will also plot the corresponding evolution of spatial patterns as obtained with the steady punishment model, which will enable us to distinguish and understand better the origins of the enhanced promotion of public cooperation.

\subsection{Mechanisms of cooperation promotion}

In agreement with the definition of the model incorporating adaptive punishment, it is important to note that cooperators with non-zero punishing activity ($\pi_x>0$) can exist only along the borders separating the C and D domains. The pure C phase, as well as interiors of large cooperative domains, lack cooperators having $\pi_x>0$ because there is a constant drift towards non-punishment if there are no D $\to$ C invasions occurring in the neighborhood.

\begin{figure}
\centerline{\epsfig{file=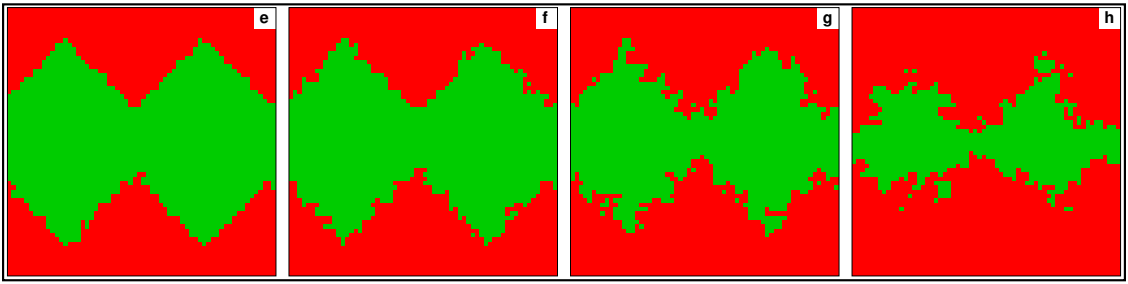,width=15cm}}
\caption{Recovery and preservation of smooth interfaces separating cooperators and defectors invigorates spatial reciprocity. Presented are characteristic snapshots of the spatial grid over time where cooperators are depicted green and defectors are depicted red. Darker shades of green indicate a higher punishing activity, while non-punishing cooperators are depicted bright green. A prepared initial state, corresponding to a rough interface, is used to reveal the difference in the impact of the two punishing models. Panels (a), (b), (c) and (d) show the evolution of the adaptive punishment model at $1$, $3$, $10$ and $500$ Monte Carlo steps, respectively. Panels (e), (f), (g) and (h), on the other hand, show the evolution of the steady punishment model at $2$, $5$, $10$ and $20$ Monte Carlo steps. Note that all the punishing cooperators in panels (e), (f), (g) and (h) are depicted with a slightly darker shade of green, as representative for players having punishing activity $\pi_x=1$. The parameters used for both models are $r=2.8$, $\alpha=1$, $\Delta=7.4$ and $L=62$ (a smaller system size was used solely for the production of these snapshots to ensure that the different shades of green in the top row are distinguishable). The final state in the adaptive punishment model is a pure C phase, while the steady punishment model yields a pure D phase (both not shown).}
\label{interface}
\end{figure}

We therefore first focus on the evolutionary dynamics along the interfaces separating the C and D domains. Note that previous studies emphasized that smooth interfaces between the competing strategies are beneficial for cooperators because it allows the network reciprocity to take full effect. On the contrary, rough interfaces provide ample opportunities for defectors to invade and spread, even at relatively high synergy factors. To demonstrate the positive effect of adaptive punishment, we start the simulation from a prepared initial state corresponding to a rough interface, as depicted in Fig.~\ref{interface}a (adaptive punishment) and Fig.~\ref{interface}e (steady punishment). By following the snapshots in the lower row (panels e, f, g and h) from left to right, we can observe that steady punishment efforts fail to restore the broken phalanx of cooperators \cite{sigmund_10}. Due to the additional roughening of the interfaces defectors can invade the cooperative domain very effectively, eventually leading to a pure D phase (not shown). Note that for clarity we have in this case incorporated punishing cooperators only, as indicated by darker green, but not pure cooperators, thus artificially excluding the emergence of negative consequences of second-order free-riding. Despite of this lenient predisposition, however, steady punishment fails for the considered parameter values.

\begin{figure}
\centerline{\epsfig{file=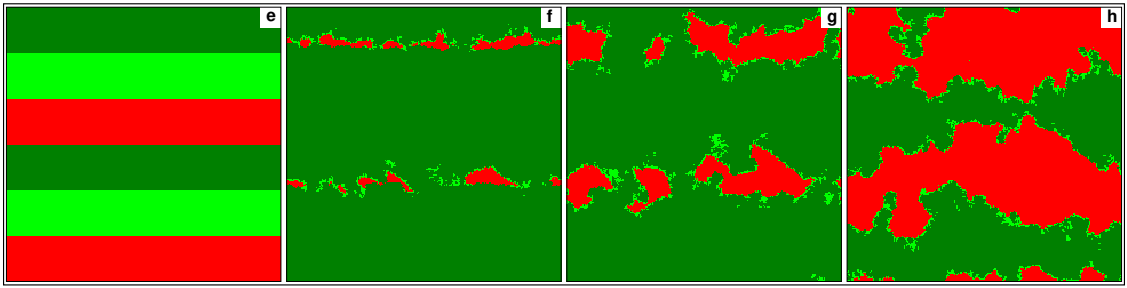,width=15cm}}
\caption{Adaptive punishment prevents the emergence of cyclic dominance. Presented are characteristic snapshots of the spatial grid over time using the same color scheme as in Fig.~\ref{interface}. Panels (e), (f), (g) and (h) show the evolution of the steady punishment model at $0$, $500$, $1000$ and $1800$ Monte Carlo steps. In this case a prepared initial state, including layers of defectors (red), punishing cooperators (dark green) and non-punishing cooperators (bright green), was necessary to reach the evolutionary stable cyclic dominant state at such a small system size. Note that punishing cooperators are able to invade defectors, defectors are able to invade non-punishing cooperators, and non-punishing cooperators are able to exploit punishing cooperators. Panels (a), (b), (c) and (d), on the other hand, show the evolution of the adaptive punishment model at identical Monte Carlo steps. To facilitate comparisons with the output of the steady punishment model, we have again used a layered initial state of cooperators and defectors. Importantly, here the punishing cooperators that emerge spontaneously along the interfaces prevent defectors of successfully invading the population. The parameters used for both models are $r=1.7$, $\alpha=0.5$, $\Delta=5$ and $L=240$.}
\label{cyclic}
\end{figure}

The impact of adaptive punishment is significantly different, as can be observed by following the snapshots in the upper row (panels a, b, c and d) from left to right. Here the localized and temporarily very active punishers, which emerge spontaneously as a response to the D $\to$ C invasions, succeed in restoring a smooth (straight) interface between cooperators and defectors. This in turn disables defectors to invade, and in fact invigorates the traditional spatial reciprocity mechanism \cite{nowak_n92b}. The demonstrated recovery and preservation of smoothness along the interfaces is the first key advantage warranted by adaptive punishment. Importantly, once the regularity of the interfaces is re-established the enhanced effectiveness of spatial reciprocity spontaneously introduces a decrease in the punishing activity, before eventually the latter altogether seizes once the pure C phase is reached. Adaptive punishment thus allows for a spontaneous but prompt and determined response to a crisis, \textit{e.g} a threatening invasion of defectors, which is unattainable with previous models.

\begin{figure}
\centerline{\epsfig{file=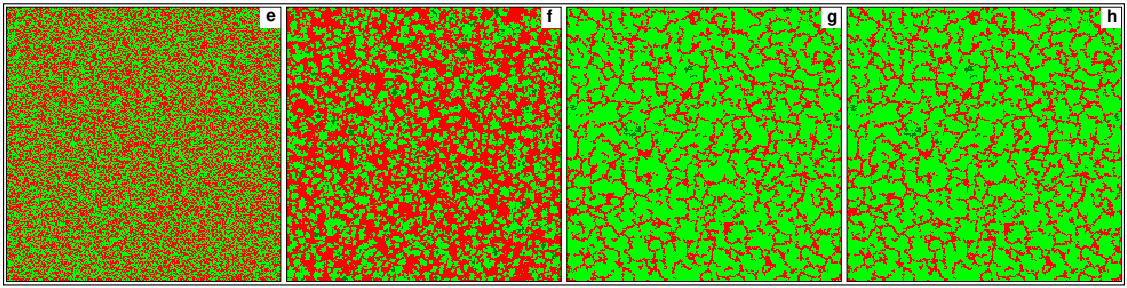,width=15cm}}
\caption{Competitive advantage of adaptive punishment in cooperation-prone environments. Presented are characteristic snapshots of the spatial grid using the same color scheme as in Fig.~\ref{cyclic}. Panels (a), (b), (c) and (d) show the evolution of the adaptive punishment model at $0$, $100$, $2000$ and $6000$ Monte Carlo steps, respectively. Although initially defectors seem to be heading towards undisputed dominance, the occurrence of a sufficiently large cooperative cluster, supported by spontaneous adaptive punishment along its borders, is able to revert the evolutionary process in favor of cooperation. Panels (e), (f), (g) and (h), on the other hand, show the evolution of the steady punishment model at $0$, $10$, $200$ and $6000$ Monte Carlo steps. In this case punishing cooperators die out fast, leaving behind a stable mixed phase of second-order free-riders (non-punishing cooperators) and defectors. The parameters used for both models are $r=4.5$, $\alpha=2$, $\Delta=1.8$ and $L=240$.}
\label{compete}
\end{figure}

The spontaneous emergence of cyclic dominance is also common in evolutionary games on structured populations \cite{szolnoki_epl10,szolnoki_pre11}. While it can be argued that cyclic dominance hinders the extinction of strategies and thus in a way ensures the survivability of cooperators, in models incorporating punishment it is far more common that the ones actually benefiting most from it are the defectors. The series of snapshots in the bottom row of Fig.~\ref{cyclic} clearly demonstrates such a scenario. Although initially the steady sanctioning efforts of punishing cooperators (dark green) seem to suffice for eliminating defectors (red), the fact that the non-punishing cooperators (bright green) are superior to those that do punish provides an escape hatch for defectors. Hence, while the second-order free-riders are superior to punishers, the punishers are superior to defectors, who are in turn superior to second-order free-riders, and so the loop of dominance becomes closed. It is important to note that prepared initial states are helpful, but certainly not necessary, for illustrating this behavior. The stability of cyclic dominance can also be confirmed from a random initial state, but for finite size effects to be avoided the system size must be as large $7000 \times 7000$ for the parameter values used in this case. Conversely, prepared, yet still non-preferential initial states, may decrease the required computational efforts by an order of magnitude. Turning to the outcome of the model with adaptive punishment, depicted in the top row of Fig.~\ref{cyclic}, it can be observed that cooperators prevail over defectors without the emergence of cyclic dominance that could render the extinction of defectors impossible. Here the adaptive nature of punishment prevents defectors from effectively exploiting cooperators, both the ones that punish as well as the ones that do not. The ones that do punish are actually superior to defectors (see bottom row), while the ones that do not are protected by smooth interfaces that maximize the effect of spatial reciprocity. Altogether, this solidifies yet another way by means of which adaptive punishment is superior to steady sanctioning efforts.

Finally, it is interesting to examine the evolutionary dynamics in situations that inherently favor the evolution of cooperation, \textit{i.e.} when the synergetic effects of collaborative efforts are strong. As above, we present in Fig.~\ref{compete} two rows of snapshots, the upper row depicting results obtained with the adaptive punishment models and the bottom row depicting results obtained with the steady punishment model. If cooperators can survive in the absence of punishment, it is straightforward to conclude that punishing cooperators (dark green) will die out over second-order free-riders (bright green), as long as the punishment is sufficiently costly \cite{helbing_ploscb10}. This is indeed what is illustrated in the bottom row of Fig.~\ref{compete}. The stationary state is therefore a mixed C+D phase, being that the synergetic effects of cooperation are not strong enough for defectors to die out completely. For the same set of parameters the adaptive punishment model yields a rather different output. The end result is a pure C phase, indicating that in welfare societies the coexistence with defectors is not necessary if individuals are temporary able to resort to punishing antisocial behavior. The fact that the punishing activity depends on the local success of both strategies enables the society to reap all the benefits of sanctioning (which ultimately is a defection-free environment), while at the same time imposes sufficiently small personal losses on those that execute it. In this way, by means of self-organization, adaptive punishment grants a competitive advantage to those that are prepared to sanction defectors in a globally economic way.

\section{Summary} \label{Discussion}
Here we have shown how adaptive punishment, triggered simply by the local success of defectors on the square lattice, is able to promote the evolution of cooperation in a society facing public goods games. Crucial thereby is the spontaneous emergence of gatekeepers, which are rare (localized) punishers that reside at the interfaces separating the domains of the two competing strategies, who are willing to punish hard those defectors that try to invade the cooperative domain. This keeps the interface between cooperators and defectors smooth, which in turn supports spatial reciprocity as a potent promoter of collective efforts even when the synergetic effects of cooperation are minute. In other parameter regions, typically when the fines are moderate but the costs substantial, adaptive punishment can destroy strategic coexistence maintained by cycling competition, and in so doing eliminates the survivability of defectors that is otherwise warranted by local spatial patterns. Finally, adaptive punishers may benefit from enhanced competitiveness, which allows them to completely eliminate defectors.

Compared to steady sanctioning efforts assumed in the majority of previously published works concerning sanctioning on structured populations \cite{brandt_prsb03,nakamaru_eer05,helbing_ploscb10}, adaptive punishment may be evolutionary advantageous in that not only it can maintain high levels of cooperation under very unfavorable conditions, but also in that the footprint of sanctioning in terms of expenses of the whole population is negligible. Adaptive punishment is thus very efficient, which fits nicely to recent results obtained on sanctioning in well-mixed populations, where it has been reported that coordinated and rare punishing efforts can be very much effective in raising group-average payoffs \cite{boyd_s10}. Given the fact that punishment is costly, it is far from obvious that the investments in sanctioning will be reimbursed by potentially higher levels of cooperation. From the results presented in this paper it is inspiring to learn that spatial structure allows for self-organization and adaptation of punishing efforts in a way that optimizes their effectiveness without explicitly modeling for this in the makeup of the game itself. Counterintuitively, a constant drift towards a non-punishing state may help to elevate social welfare by minimizing the disadvantages of a large-scale sanctioning effort, while at the same time practically nothing has to be sacrificed in terms of effectiveness of deterring antisocial behavior.

Over-fishing, environmental pollution, depletion of natural resources, or the misuse of social security systems, are prime examples of the exploitation of public goods, and it is vital that we identify the most effective mechanisms that prevent such counterproductive behavior. We believe that game theoretical models incorporating spatial structure \cite{szabo_pr07}, although adding to the complexity of solutions especially when containing more than two strategies, represent a viable route to achieving this goal. Results presented in this paper indicate that punishment should not be dismissed and promote further quantitative research in social systems with methods from statistical physics \cite{castellano_rmp09}.

\ack
This research was supported by the Slovenian Research Agency (grant J1-4055) and the Hungarian National Research Fund (grant K-73449).

\section*{References}
\providecommand{\newblock}{}

\end{document}